\documentclass[preprint,showpacs,preprintnumbers,amsmath,amssymb]{revtex4-1}
\usepackage{epsfig}
\usepackage{graphicx}
\usepackage{dcolumn}
\usepackage{bm}

\newcommand{\ord}{{\cal O}}
\def\beq{\begin{equation}}
\def\eeq#1{\label{#1}\end{equation}}
\def\eeqn{\end{equation}}
\newcommand\iden{\leavevmode\hbox{\small1\normalsize\kern-.33em1}}

\newcommand{\SLASH}[2]{\makebox[#2ex][l]{$#1$}/}
\newcommand{\pslash}{\SLASH{p}{.2}}
\newcommand{\sq}{\sqrt{2}}
\newcommand{\bea} {\begin{eqnarray}}
\newcommand{\eea} {\end{eqnarray}}

\newcommand{\Lg}{{\mathcal L}}
\newcommand{\rd}{\partial}

\newcommand{\sbt}{s_{\beta}}
\newcommand{\cbt}{c_{\beta}}
\newcommand{\tbt}{t_{\beta}}
\newcommand{\sbcb}{s_{\beta} c_\beta}
\newcommand{\ea}{\epsilon_{1}}
\newcommand{\eb}{\epsilon_{2}^{*}}

\let\jnfont=\rm
\def\NPB#1,{{\jnfont Nucl.\ Phys.\ B }{\bf #1},}
\def\PLB#1,{{\jnfont Phys.\ Lett.\ B }{\bf #1},}
\def\EPJC#1,{{\jnfont Eur.\ Phys.\ Jour.\ C }{\bf #1},}
\def\PRD#1,{{\jnfont Phys.\ Rev.\ D }{\bf #1},}
\def\PRL#1,{{\jnfont Phys.\ Rev.\ Lett.\ }{\bf #1},}
\def\MPLA#1,{{\jnfont Mod.\ Phys.\ Lett.\ A }{\bf #1},}
\def\JPG#1,{{\jnfont J.\ Phys.\ G }{\bf #1},}
\def\CTP#1,{{\jnfont Commun.\ Theor.\ Phys.\ }{\bf #1},}
\def\JHEP#1,{{\jnfont JHEP \ }{\bf #1},}
\def\NPPS#1,{{\jnfont Nucl.\ Phys.\ Proc.\ Suppl.\ }{\bf #1},}
\def\CPC#1,{{\jnfont Computl.\ Phys.\ Commun.\ }{\bf #1},}
\def\APPB#1,{{\jnfont Acta\ Phys.\ Polon.\ B }{\bf #1},}

\def\EPL#1,{{\jnfont Europhys.\ Lett. }{\bf #1},}

\begin{document}

\title{\ \\[10mm] Rare Z-decay into light pseudoscalar bosons \\ in the simplest little Higgs model}

\author{Lei Wang, Xiao-Fang Han}

\affiliation{ Department of Physics, Yantai University, Yantai
264005, China}

\begin{abstract}
The simplest little Higgs model predicts a light pseudoscalar boson
$\eta$ and opens up some new decay modes for $Z$-boson, such as $Z
\to \bar{f} f \eta$, $Z\to \eta\eta\eta$, $Z\to \eta\gamma$ and
$Z\to \eta gg$. We examine these decay modes in the parameter space
allowed by current experiments, and find that the branching ratios
can reach $10^{-7}$ for $Z\to \bar{b}b\eta$, $10^{-8}$ for $Z\to
\bar{\tau}\tau\eta$, and $10^{-8}$ for $Z\to \eta\gamma$, which
should be accessible at the GigaZ option of the ILC. However, the
branching ratios can reach $10^{-12}$ for $Z\to \eta\eta\eta$, and
$10^{-11}$ for $Z\to \eta gg$, which are hardly accessible at the
GigaZ option.

\end{abstract}


\maketitle

\section{Introduction}
Little Higgs theory \cite{LH} has been proposed as an interesting
solution to the hierarchy problem. So far various realizations of
the little Higgs symmetry structure have been proposed
\cite{otherlh,lht,lst,sst}, which can be categorized generally into
two classes \cite{smoking}. One class use the product group,
represented by the littlest Higgs model \cite{lst}, in which the SM
$SU(2)_L$ gauge group is from the diagonal breaking of two (or more)
gauge groups. The other class use the simple group, represented by
the simplest little Higgs model (SLHM) \cite{sst}, in which a single
larger gauge group is broken down to the SM $SU(2)_L$.

Since these little Higgs models mainly alter the properties of the
Higgs boson, hints of these models may be unraveled from various
Higgs boson processes. The phenomenology of Higgs boson in these
little Higgs models has been widely studied
\cite{hrrhan,higgslh,slhv,wketa}. In addition to the SM-like Higgs
boson $h$, the SLHM predicts a pseudoscalar boson $\eta$, whose mass
can be as low as $\ord{(10~\mathrm{GeV})}$. The constraint from the
non-observation in the decay $\Upsilon\to \gamma \eta$ excludes
$\eta$ with mass below 5-7 GeV \cite{5-7GeV}. Thus, the $Z$-decays
into $\eta$ can be open for a light $\eta$, such as $Z \to \bar{f} f
\eta$ ($f=b,\tau$), $Z\to \eta\eta\eta$, $Z\to \eta\gamma$ and $Z\to
\eta gg$.

The next generation $Z$ factory can be realized in the GigaZ option
of the International Linear Collider (ILC)\cite{gigaz}. The ILC is a
proposed electron-positron collider with tunable energy ranging from
400GeV to 500GeV and polarized beams in its first phase, and the
GigaZ option corresponds to its operation on top of the resonance of
$Z$-boson by adding a bypass to its main beam line. About $2 \times
10^9$ $Z$ events can be generated in an operational year of $10^7 s$
of GigaZ, which implies that the expected sensitivity to the
branching ratio of $Z$-decay can be improved from $10^{-5}$ at the
LEP to $10^{-8}$ at the GigaZ \cite{gigaz}. Therefore, it will offer
an important opportunity to probe the new physics via the exotic or
rare decays of $Z$-boson. The $Z$-boson flavor-changing
neutral-current (FCNC) decays have been studied in many new physics
models. For the lepton (quark) flavor violation decays, the
branching ratios can be enhanced sizably compared with the SM
predictions \cite{lep,quark}. In addition to the FCNC processes, the
branching ratios for the processes of the $Z$-decay into light Higgs
boson(s) in some new physics models can be accessible at the GigaZ
option of the ILC \cite{cao,toploop,lighthiggs}. In this paper, we
will focus on the processes of the $Z$-decays into $\eta$ in the
SLHM, namely $Z \to \bar{f} f \eta$ ($f=b,\tau$), $Z\to
\eta\eta\eta$, $Z\to \eta\gamma$ and $Z\to \eta gg$.

This work is organized as follows. In Sec. II we recapitulate the
SLHM. In Sec. III we study the five decay modes, respectively.
Finally, we give our conclusion in Sec. IV.

\section{Simplest little Higgs model}
The SLHM is based on $[SU(3) \times U(1)_X]^2$ global symmetry. The
gauge symmetry $SU(3) \times U(1)_X$ is broken down to the SM
electroweak gauge group by two copies of scalar fields $\Phi_1$ and
$\Phi_2$, which are triplets under the $SU(3)$ with aligned VEVs
$f_1$ and $f_2$. The uneaten five pseudo-Goldstone bosons can be
parameterized as

\beq
\Phi_{1}= e^{ i\; t_\beta \Theta } \left(\begin{array}{c} 0 \\
0 \\ f_1
\end{array}\right)\;,\;\;\;\;
\Phi_{2}= e^{- \; \frac{i}{t_\beta} \Theta} \left(\begin{array}{c} 0 \\  0 \\
f_2
\end{array}\right)\;,
\label{paramet}
\end{equation}
where
\begin{equation}
   \Theta = \frac{1}{f} \left[
        \left( \begin{array}{cc}
        \begin{array}{cc} 0 & 0 \\ 0 & 0 \end{array}
            & H \\
        H^{\dagger} & 0 \end{array} \right)
        + \frac{\eta}{\sqrt{2}}
        \left( \begin{array}{ccr}
        1 & 0 & 0 \\
        0 & 1 & 0 \\
        0 & 0 & 1 \end{array} \right) \right],
\end{equation}
$f=\sqrt{f_1^2+f_2^2}$ and $t_\beta\equiv \tan\beta= f_2 / f_1$.
Under the $SU(2)_L$ SM gauge group, $\eta$ is a real scalar, while
$H$ transforms as a doublet and can be identified as the SM Higgs
doublet. The kinetic term in the non-linear sigma model is \beq
\label{eq:Lg:gauge0} \Lg_\Phi = \sum_{j=1,2}\left| \left(\rd_\mu + i
g A^a_\mu T^a - i \frac{g_x}{3} B^x_\mu \right) \Phi_j \right|^2,
\end{equation}
where $g_x =g\tan \theta_W/ \sqrt{1-\tan^2 \theta_W/3}$ with
$\theta_W$ being the electroweak mixing angle. As $\Phi_1$ and
$\Phi_2$ develop their VEVs, the new heavy gauge bosons $Z'$, $Y^0$,
and $X^{\pm}$ get their masses proportional to $f$. A novel coupling
of the SM-like Higgs boson $h$ can be derived from the Eq.
(\ref{eq:Lg:gauge0}), \beq \label{zhy} \Lg_{Zh\eta}
=\frac{m_Z}{\sqrt{2}f}(t_\beta-\frac{1}{t_\beta})
(\eta\partial^{\mu}h-h\partial^{\mu}\eta).\end{equation}

The gauged $SU(3)$ symmetry promotes the SM fermion doublets into
$SU(3)$ triplets. There are two possible gauge charge assignments
for the fermions: the 'universal' embedding and the 'anomaly-free'
embedding.  The first choice is not favored by the electroweak
precision data \cite{sst}, so we focus on the second way of
embedding. The quark Yukawa interactions for the third generation
and the first two generations can be written respectively as\bea
{\cal L}_3 &=& i \lambda_1^t t_1^c \Phi_1^{\dagger} Q_3
  + i \lambda_2^t t_2^c \Phi_2^{\dagger} Q_3
  + i  \frac{\lambda_d^m}{\Lambda}  d_m^c \epsilon_{ijk}
      \Phi_1^i \Phi_2^j Q_3^k + h.c., \label{simtopyukawa}\\
{\cal L}_{1,2} &=&  i \lambda_1^{d_n} d_{1n}^c Q_n^{T} \Phi_1
  + i \lambda_2^{d_n} d_{2n}^c Q^{T}_n \Phi_2
  + i \frac{\lambda_{u}^{mn}}{\Lambda} u_m^c \epsilon_{ijk} \Phi_1^{*i}
    \Phi_2^{*j} Q_n^k + h.c.,\label{simucyukawa}
\eea where $n=1,2$ are the first two generation indices;
$i,j,k=1,2,3$; $Q_3=\{ t_L, b_L, i T_L\}$ and $Q_n = \{ d_{nL}, -
u_{nL}, i D_{nL}\}$; $d_m^c$ runs over $(d^c, s^c, b^c, D^c, S^c)$;
$d^c_{1n}$ and $d^c_{2n}$ are linear combinations of $d^c$ and $D^c$
for $n=1$ and of $s^c$ and $S^c$ for $n=2$; $u^c_m$ runs over $(u^c,
c^c, t^c, T^c)$. For simplicity, we assume the quark flavor mixing
are small and neglect the mixing effects. Eqs. (\ref{simtopyukawa})
and (\ref{simucyukawa}) contain the Higgs boson interactions and the
mass terms for the three generations of quarks: \bea
\label{tTmixing} {\cal L}_t &\simeq&-f \lambda_2^t \left[
x_\lambda^t c_\beta  t_1^c(-s_1t_L
   +c_1T_L)G_1(\eta)+s_\beta t_2^c (s_2 t_L+ c_2 T_L) G_2(\eta) \right]+h.c.,\,\\
   \label{dDmixing}
{\cal L}_{d_n} &\simeq&-f \lambda_2^{d_n} \left[ x_\lambda^{d_n} c_\beta d_1^c
  (s_1 d_{nL}+c_1 D_{nL})G^*_1(\eta)+s_\beta d_2^c (-s_2 d_{nL}+c_2 D_{nL})G^*_2(\eta)\right]+h.c.,\,\\
{\cal L}_{b} &\simeq&-\frac{\lambda_b}{\Lambda}f^2 s_\beta c_\beta
s_3 b^c b_{_L}G_3(\eta)
          +h.c.,\,\label{lgb}\\
{\cal L}_{q} &\simeq&-\frac{\lambda_q}{\Lambda}f^2 s_\beta c_\beta
s_3 q^c q_{_L}G^*_3(\eta)+h.c. \ (q=u,c),\, \label{lgq}\eea where
 \bea x_\lambda^t&\equiv & {\lambda_1^t \over \lambda_2^t},\
\ x_\lambda^{d_n}\equiv {\lambda_1^{d_n} \over \lambda_2^{d_n}},\ \
s_{\beta}\equiv\frac{f_2}{\sqrt{f^2_1+f^2_2}},\ \
c_{\beta}\equiv\frac{f_1}{\sqrt{f^2_1+f^2_2}},\ \ \nonumber\\
s_1&\equiv & \sin {t_\beta (h+v)\over \sqrt{2}f},\ \ s_2\equiv
\sin{(h+v) \over \sqrt{2}t_\beta f},\ \ s_3\equiv
\sin{(h+v)(t_\beta^2+1)\over \sqrt{2}t_\beta f},\ \ \nonumber\\
G_1(\eta)&\equiv &
1-i\frac{t_{\beta}}{\sqrt{2}f}\eta-\frac{t_{\beta}^2}{4f^2}\eta^2,\
\
G_2(\eta)\equiv 1+i\frac{1}{\sqrt{2}t_{\beta}f}\eta-\frac{1}{4t_{\beta}^2f^2}\eta^2,\ \ \nonumber\\
G_3(\eta)&\equiv &
1+i\frac{1}{\sqrt{2}f}(t_\beta-\frac{1}{t_\beta})\eta
-\frac{1}{4f^2}(t_\beta-\frac{1}{t_\beta})^2\eta^2, \eea with $h$
and $v$ being the SM-like Higgs boson field and its VEV,
respectively. The mass eigenstates are obtained by mixing the
corresponding interaction eigenstates, e.g., the mass eigenstates
$(t_{mL}, T_{mL})$ and $(t_m^c, T_m^c)$ are  respectively the
mixtures of $(t_{L}, T_{L})$ and $(t^c, T^c)$. The diagonalization
of the mass matrix in Eqs. (\ref{tTmixing}) and (\ref{dDmixing}) is
performed numerically in our analysis, and the top quark
$(T,~D,~d,~S,~s)$ couplings of $h$ and $\eta$ bosons can also be
obtained without resort to any expansion of $v/f$. From Eqs.
(\ref{lgb}) and (\ref{lgq}), we can get directly the couplings
$-i\frac{m_f}{v}x_{f} h\bar{f}f$ and $\frac{m_f}{v}y_{_f}
\eta\bar{f}\gamma_{5}f$ $(f=b,u,c)$ with \beq
x_f=\frac{v}{\sqrt{2}f}(t_\beta+\frac{1}{t_\beta})\cot\frac{v}{\sqrt{2}f}(t_\beta+\frac{1}{t_\beta}),
\ \ \ y_u=y_c=-y_b=\frac{v}{\sqrt{2}f}(t_\beta-\frac{1}{t_\beta}).
\end{equation}
The charged lepton couplings with $h$ and $\eta$ are the same as
those of $b$ quark, but replacing $m_b$ with the corresponding
lepton mass. Hereafter we denote the mass eigenstates without the
subscript '$m$' for simplicity.

The Yukawa and gauge interactions break the global symmetry and then
provide a potential for the Higgs boson. However, the
Coleman-Weinberg potential alone is not sufficient since the
generated $h$ mass is too heavy and the new pseudoscalar $\eta$ is
massless. Therefore, one can introduce a tree-level $\mu$ term which
can partially cancel the $h$ mass \cite{sst,slhv}: \beq -\mu^2
(\Phi^\dagger_1 \Phi_2 + h.c.) = - 2 \mu^2 f^2 \sbt\cbt \cos\left(
\frac{\eta}{\sq \sbt\cbt f} \right)
 \cos \left(
 \frac{\sqrt{H^\dagger H}}{f \cbt\sbt}
\right).
\end{equation}
The Higgs potential becomes \beq \label{eq:VCW} V = - m^2 H^\dagger
H + \lambda (H^\dagger H)^2
 - \frac{1}{2} m_\eta^2 \eta^2 +\lambda' H^\dagger H \eta^2 + \cdots,
\end{equation}
where \beq \label{eq:msq:lambda} m^2 = m_0^2 - \frac{\mu^2}{\sbcb},
\quad \lambda =\lambda_0 - \frac{\mu^2}{12\sbt^3 \cbt^3f^2}, \quad
\lambda' = - \frac{\mu^2}{4 f^2 \sbt^3 \cbt^3}
\end{equation}
with $m_0$ and $\lambda_0$ being respectively the one-loop
contributions to the $H$ mass and the quartic couplings from the
contributions of fermion loops and gauge boson loops \cite{sst}.

After EWSB, the coupling $h\eta\eta$ can be obtained from the term
$\lambda' H^\dagger H \eta^2$ in Eq. (\ref{eq:VCW}). The Higgs VEV
and the masses of $h$ and $\eta$ are given by \beq
\label{eq:vsq:mH:meta} v^2 = \frac{ m^2}{\lambda} , \quad m_h^2 = 2
m^2 , \quad m_\eta^2 = \frac{\mu^2}{\sbcb} \cos\left(
\frac{v}{\sqrt{2} f \sbcb} \right).
\end{equation} The Coleman-Weinberg potential
involves the following parameters: \beq \label{para} f,~
x_\lambda^t,~ t_\beta,~\mu,~m_\eta,~m_h,v.
\end{equation}
Due to the modification of the observed $W$ gauge boson mass, $v$ is
defined as \cite{slhv} \beq \label{eq:v} v \simeq v_0 \left[ 1+
\frac{v_0^2}{12 f^2}\frac{\tbt^4-\tbt^2+1}{\tbt^2} -
\frac{v_0^4}{180 f^4}\frac{\tbt^8-\tbt^6+\tbt^4-\tbt^2+1}{\tbt^4}
\right],
\end{equation}
where $v_0=246$ GeV is the SM Higgs VEV. Assuming that there are no
large direct contributions to the potential from physics at the
cutoff, we can determine other parameters in Eq. (\ref{para}) from
$f$, $t_\beta$ and $m_{\eta}$ with the definition of $v$ in Eq.
(\ref{eq:v}).

\section{the rare Z-decay into $\eta$}
\begin{figure}[tb]
\begin{center}
 \epsfig{file=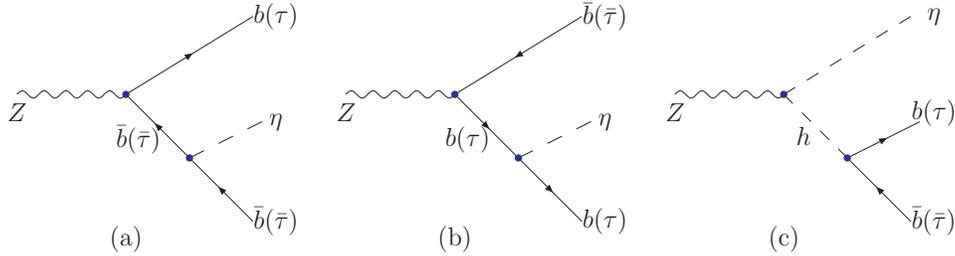,width=12.8cm}
\end{center}
\vspace{-0.6cm} \caption{Feynman diagrams for $Z\to \bar{f}f\eta$
$(f=b,~\tau)$ in the SLHM.} \label{zffy}
\end{figure}

\begin{figure}[tb]
\begin{center}
 \epsfig{file=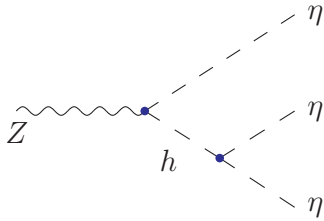,width=4.5cm}
\end{center}
\vspace{-1.4cm} \caption{Feynman diagrams for $Z\to \eta\eta\eta$ in
the SLHM. } \label{zyyy}
\end{figure}

\begin{figure}[tb]
\begin{center}
 \epsfig{file=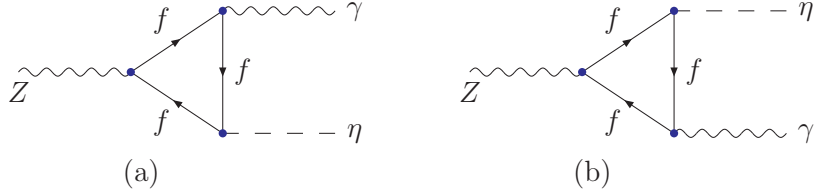,width=11cm}
\end{center}
\vspace{-1.0cm} \caption{Feynman diagrams for $Z\to \eta\gamma$ in
the SLHM. $f$ denotes the charged fermions in SM, the new quarks
$T$, $D$, and $S$.} \label{zry}
\end{figure}

\begin{figure}[tb]
\begin{center}
 \epsfig{file=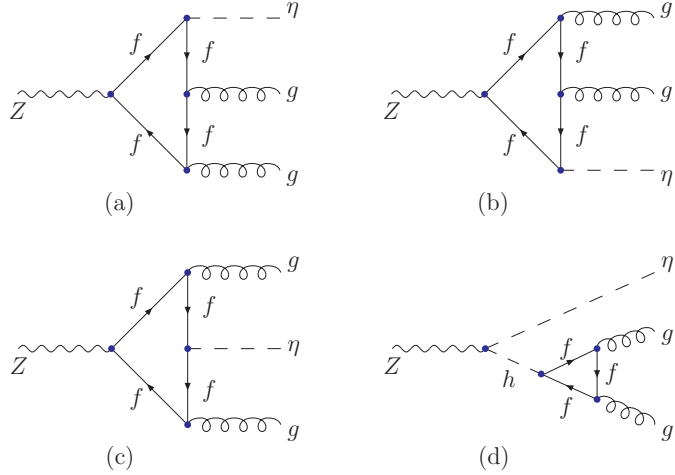,width=9cm}
\end{center}
\vspace{-1.0cm} \caption{Feynman diagrams for $Z\to \eta gg$ in the
SLHM. $f$ denotes the SM quarks, the new quarks $T$, $D$, and $S$.
The diagrams by exchanging the two gluons are not shown here.}
\label{zggy}
\end{figure}

In the SLHM, the rare $Z$-decays $Z \to \bar{f} f \eta$
($f=b,\tau$), $Z\to \eta\eta\eta$, $Z\to \eta\gamma$ and $Z\to \eta
gg$ can be depicted by the Feynman diagrams shown in Fig.
\ref{zffy}, Fig. \ref{zyyy}, Fig. \ref{zry} and Fig. \ref{zggy},
respectively. For the decay $Z\to \eta\eta\eta$, ref. \cite{cao}
shows the contributions of the scalar-loop diagrams can also be
important for the large $h\eta\eta$ coupling. Here we do not
consider the contributions of the loop diagram since the $h\eta\eta$
coupling does not have the large factor of enhancement in the SLHM
(see $\lambda'$ in Eq.(\ref{eq:msq:lambda})).

The calculations of the loop diagrams in Fig. \ref{zry} and Fig.
\ref{zggy} are straightforward. Each loop diagram is composed of
some scalar loop functions \cite{Hooft} which are calculated by
using LoopTools \cite{Hahn}. In appendix A, we list the amplitudes
for $Z \to \bar{f} f \eta$, $Z\to \eta\eta\eta$ and $Z\to \eta
\gamma$, respectively. The expressions for the amplitude of $Z\to
\eta gg$ are very lengthy, which are not presented here.


In our calculations, we take $m_t=173.3$ GeV \cite{10073178} and the
other SM input parameters as ref. \cite{pdg}. The free SLHM
parameters are $f, ~t_\beta,~ m_{\eta}, ~x_\lambda^{d}$ and
$x_\lambda^{s}$. As shown above, the parameters
$x_\lambda^t,~\mu,~m_h$ can be determined by $f$, $t_\beta$,
$m_{\eta}$ and $v$. To satisfy the bound of LEP2, we require that
$m_h$ is larger than 114.4 GeV \cite{lep2}. Certainly, due to the
suppression of $hZZ$ coupling \cite{slhv}, the LEP2 bound on $m_h$
should be loosened to some extent. The recent studies about $Z$
leptonic decay and $e^+e^-\to \tau^+\tau^-\gamma$ process at the
$Z$-pole show that the scale $f$ should be respectively larger than
5.6 TeV and 5.4 TeV \cite{f5.65.4}. Here, we assume the new flavor
mixing matrices in lepton and quark sectors are diagonal
\cite{smoking,lpv1,lpv2}, so that $f$ and $t_\beta$ are free from
the experimental constraints of the lepton and quark flavor
violating processes. The large values of $f$ can suppress the SLHM
predictions sizably. However, the factor $t_{\beta}$ in the
couplings of $h$ and $\eta$ can be taken as a large value to cancel
the suppression of $f$ partially. For the perturbation to be valid,
$t_{\beta}$ cannot be too large for a fixed $f$. If we require
$\ord(v_0^4/f^4)/\ord(v_0^2/f^2) < 0.1$ in the expansion of $v$,
$t_\beta$ should be below 28 for $f=5.6$ TeV. In our calculations,
we take $f=5.6$ TeV and $t_{\beta}=28,~26,~24$, respectively.

The small mass of the $d$($s$) quark requires one of the couplings
$\lambda^{d}_1$ and $\lambda^{d}_2$ ($\lambda^{s}_1$ and
$\lambda^{s}_2$) to be very small, so there is almost no mixing
between the SM down-type quarks and their heavy partners. We assume
$\lambda^{d}_1$($\lambda^{s}_1$) is small, and take
$x_\lambda^{d}=1.1\times 10^{-4}$ $(x_\lambda^s=2.1\times10^{-3})$,
which can make the masses of $D$ and $S$ be in the range of $1$ TeV
and $2$ TeV for other parameters taken in our calculations. In fact,
our results show that different choices of $x_\lambda^{d}$ and
$x_\lambda^{s}$ can not have sizable effects on the result.

\begin{figure}[tb]
\begin{center}
 \epsfig{file=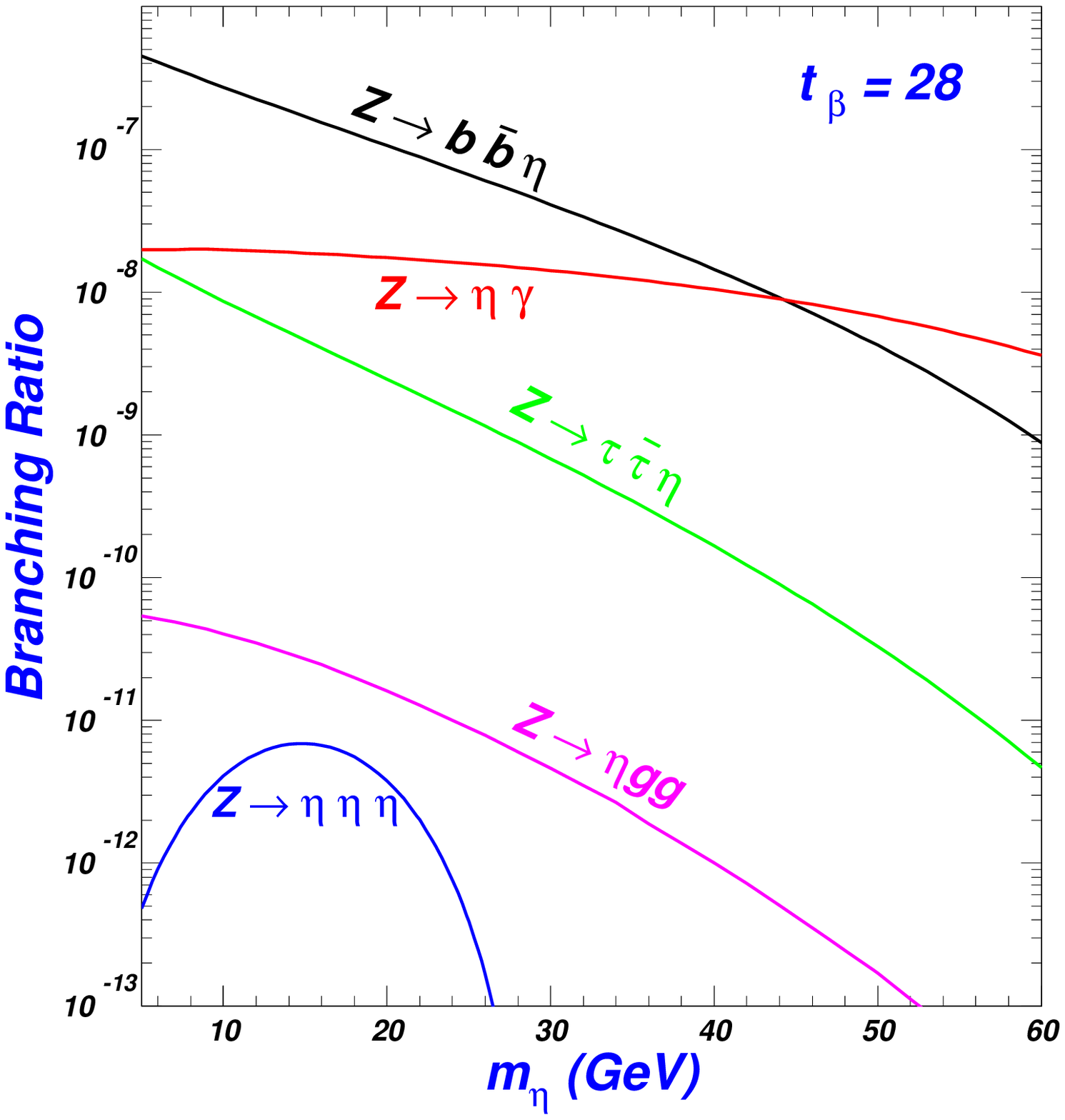,width=5.55cm}
 \epsfig{file=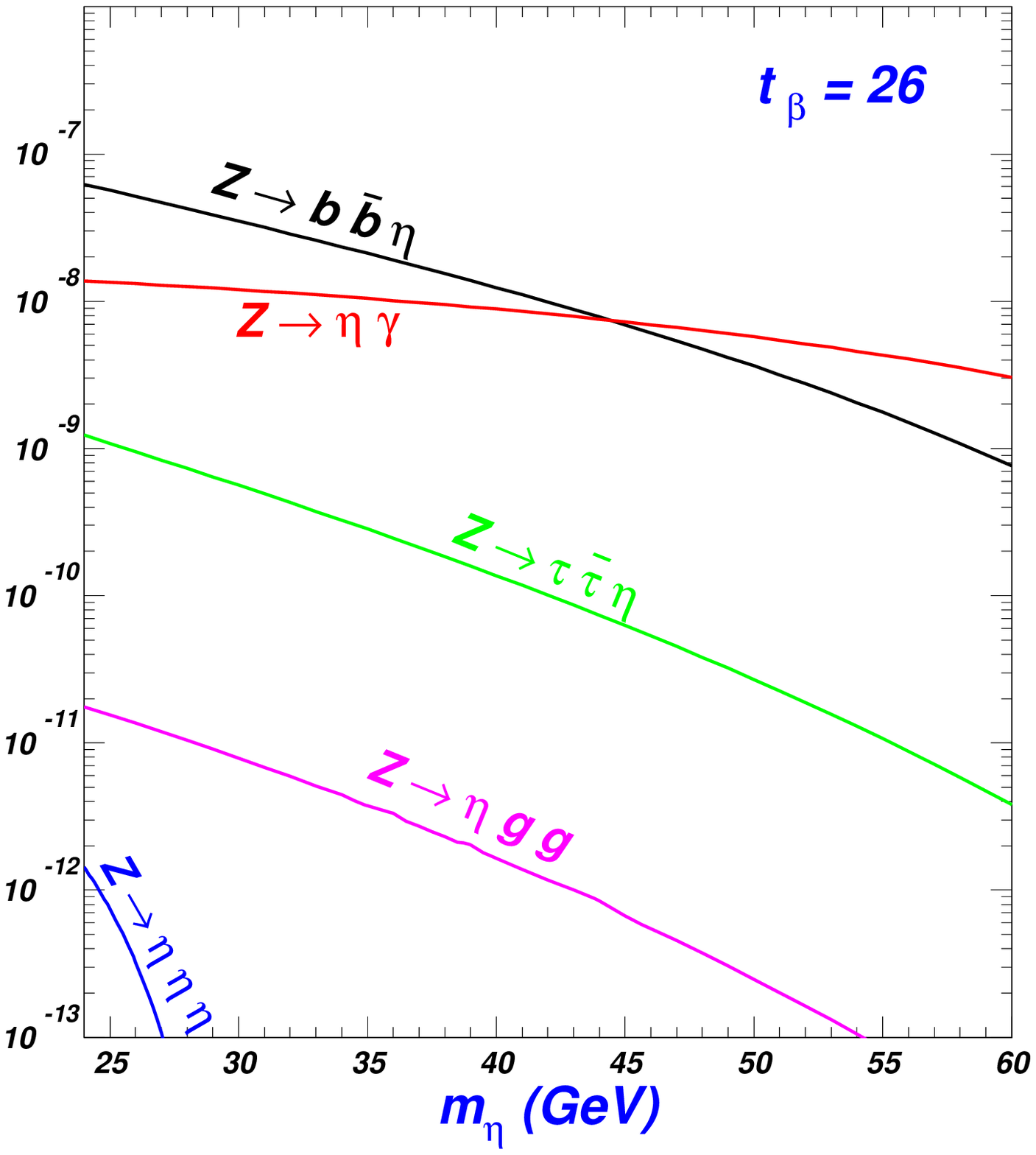,width=5.3cm}
  \epsfig{file=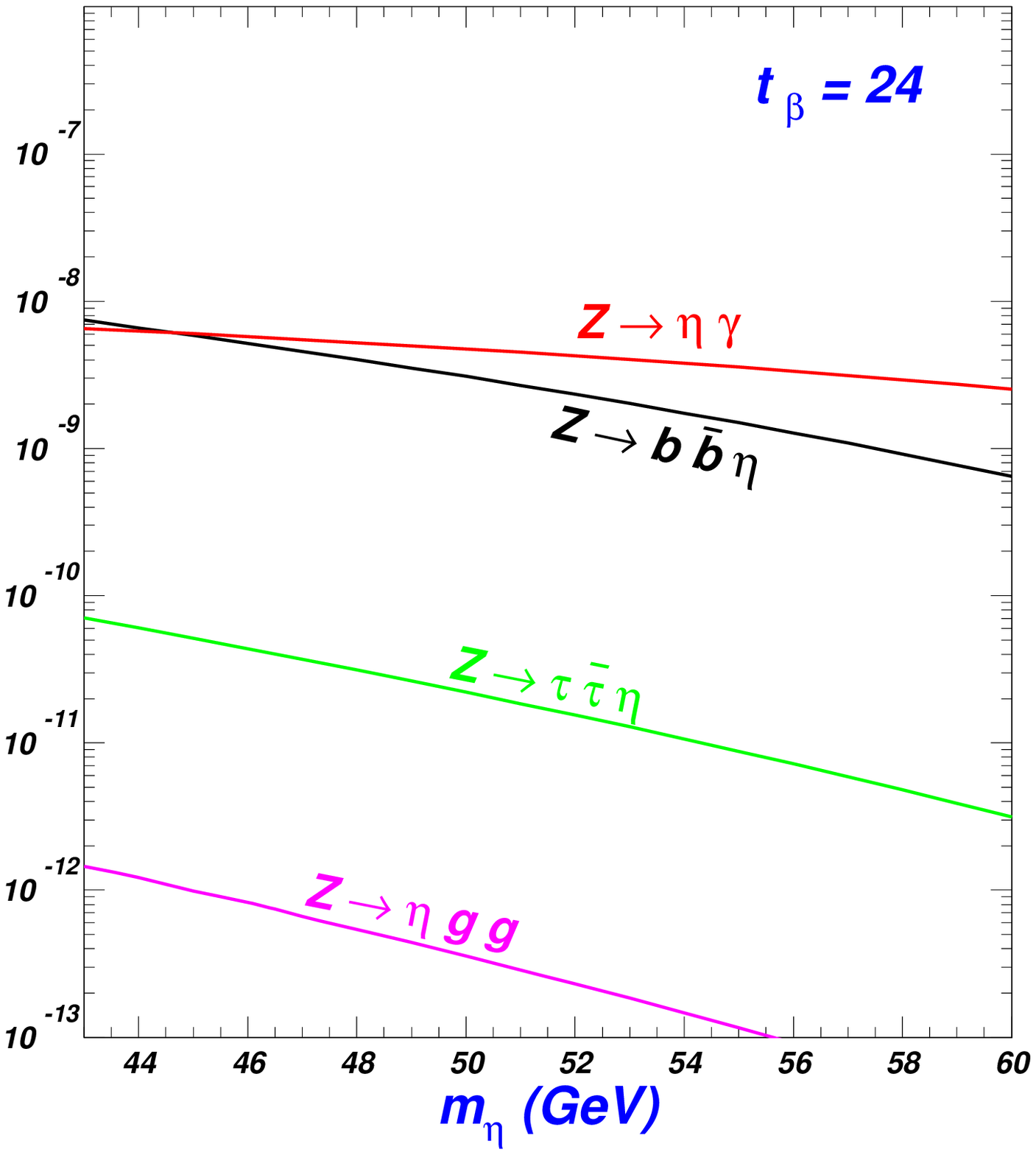,width=5.3cm}
\end{center}
\vspace{-1.0cm} \caption{For $f=5.6$ TeV, the branching ratios of $Z
\to \bar{f} f \eta$ ($f=b,\tau$), $Z\to \eta\eta\eta$, $Z\to
\eta\gamma$ and $Z\to \eta gg$ versus the $\eta$ boson mass. The
incomplete lines for $t_\beta=26$ and $24$ show the lower bounds of
the $\eta$ mass, respectively.} \label{cross}
\end{figure}

In Fig. \ref{cross}, we plot the decay branching ratios of $Z \to
\bar{f} f \eta$ ($f=b,\tau$), $Z\to \eta\eta\eta$, $Z\to \eta\gamma$
and $Z\to \eta gg$ versus the $\eta$ mass for $t_\beta=28,~26,~24$,
respectively. Because of the constraint of $m_h>114.4$ GeV, the
lower bound of $m_\eta$ is enhanced for the small $t_\beta$. Fig.
\ref{cross} shows that the ratios can reach $7\times10^{-12}$ for
$Z\to \eta\eta\eta$ with $m_\eta=15$ GeV and $t_\beta=28$, and
$5\times10^{-11}$ for $Z\to \eta gg$ with $m_\eta=5$ GeV and
$t_\beta=28$, which are too small to be detectable at the Gigaz
option of the ILC. The reason should be partly from the suppression
of three-body phase space. Besides, for the decay $Z\to \eta gg$, we
find there is a strong cancellation between the contributions of
different diagrams shown in Fig. \ref{zggy}, which can reduce the
branching ratio sizably. As the increasing of $m_\eta$, the phase
space of $Z\to \eta\eta\eta$ is suppressed, and the $h\eta\eta$
coupling which is proportional to $m^2_\eta$ is enhanced, so the
ratio reaches its peak for $m_\eta=15$ GeV and $t_\beta=28$. In
addition to the couplings $Z\eta h$ and $h\eta\eta$ which are
suppressed by the scale $f$, the factor $\frac{1}{\hat{s}-m_h^2}$
contributed by the intermediate state $h$ shown in Fig. \ref{zyyy}
(see Eq. (A4)) can suppress the branching ratio of $Z\to
\eta\eta\eta$ sizably.

Fig. \ref{cross} shows that the ratios can reach $10^{-7}$ for $Z\to
\eta\bar{b}b$, $10^{-8}$ for $Z\to \eta\bar{\tau}\tau$, and
$10^{-8}$ for $Z\to \eta\gamma$ for a light $\eta$ boson with
$t_\beta=28$, which should be accessible at the Gigaz option. The
three ratios decrease with increasing of $m_\eta$, especially for
the decays $Z\to \eta\bar{b}b$ and $Z\to \eta\bar{\tau}\tau$. For
$t_\beta=26$, the lower bound of $m_\eta$ is enhanced to 24 GeV to
satisfy the constraint of $m_h>114.4$ GeV, and the ratios of $Z\to
\eta\bar{\tau}\tau$ is below $10^{-8}$, which is hardly accessible
at the Gigaz option. For $t_\beta=24$, the lower bound of $m_\eta$
is enhanced to 43 GeV, and the ratios of the three decays are too
small to detectable at the Gigaz option. The branching ratio of
$Z\to\eta\bar{b}b$ always dominates over $Z\to\eta\bar{\tau}\tau$
since the coupling $\eta \bar{b}b$ is larger than $\eta
\bar{\tau}\tau$, and the former can be enhanced by the color factor.
Although the decay $Z\to \eta\gamma$ occurs at one-loop level, the
branching ratio can be comparable with those of $Z\to \eta\bar{f}f$
because of the large couplings $\eta \bar{t}t$ and $\eta \bar{T}T$,
and suppression of the three-body phase space for $Z\to
\eta\bar{f}f$.

\section{Conclusion}
In the framework of the simplest little Higgs model, we studied the
rare Z-decays $Z \to \bar{f} f \eta$, $Z\to \eta\eta\eta$, $Z\to
\eta\gamma$ and $Z\to \eta gg$. In the parameter space allowed by
current experiments, the branching ratios can reach $10^{-7}$ for
$Z\to \bar{b}b\eta$, $10^{-8}$ for $Z\to \bar{\tau}\tau\eta$, and
$10^{-8}$ for $Z\to \eta\gamma$, which should be accessible at the
GigaZ option of the ILC. However, the branching ratios can reach
$10^{-12}$ for $Z\to \eta\eta\eta$, and $10^{-11}$ for $Z\to \eta
gg$, which are too small to be detectable at the Gigaz.

\section*{Acknowledgment}
We thank Junjie Cao, Zhaoxia Heng, Wenlong Sang, and Jin Min Yang
for discussions. This work was supported in part by the National
Natural Science Foundation of China (NNSFC) under grant No.
11005089, and by the Foundation of Yantai University under Grant No.
WL10B24.

\appendix
\section{Amplitudes of $Z\to \bar{f}f\eta$, $Z\to \eta\eta\eta$ and $Z\to
\eta\gamma$} Here we give the amplitude of the process
$Z(p_1,\epsilon_1)\to \bar{f}(p_4)f(p_3)\eta(p_2)$ $(f=b,\tau)$. The
expressions are given by\beq M=M_a+M_b+M_c,\end{equation} where
 \bea M_a&=&-\frac{1}{(p_1-p_4)^2-m_{f}^{2}}\bar{u} (p_3)(g_{L}^{\eta}P_L+g_{R}^{\eta}P_R)
 (\pslash_1-\pslash_4 + m_f)
 \gamma^{\mu}(g_{L}^{Z}P_L+g_{R}^{Z}P_R)v(p_4)\epsilon_1(p_1),\nonumber\\
M_b&=&-\frac{1}{(p_1-p_3)^2-m_{f}^{2}}\bar{u}
(p_3)\gamma^{\mu}(g_{L}^{Z}P_L+g_{R}^{Z}P_R)
 (\pslash_3-\pslash_1 + m_f)(g_{L}^{\eta}P_L+g_{R}^{\eta}P_R)
 v(p_4)\epsilon_1(p_1),\nonumber\\
M_c&=&\frac{g_{hf\bar{f}} g_{Zh\eta}}{(p_3+p_4)^2-m_{h}^{2}}\bar{u}
(p_3) v(p_4) (p_1 - 2p_3 - 2p_4)^{\mu} \epsilon_{\mu}(p_1).
 \eea
The amplitude of $Z(p_1,\epsilon_1)\to \eta(p_2)\eta(p_3)\eta(p_4)$
is as follows: \beq
M=M_1(p_1,p_2,p_3,p_4)+M_1(p_1,p_3,p_2,p_4)+M_1(p_1,p_4,p_3,p_2),\end{equation}
where \beq M_1(p_1,p_2,p_3,p_4)=\frac{g_{h\eta\eta}
g_{Zh\eta}}{\hat{s}-m_{h}^{2}}(p_3+p_4-p_2)^{\mu}
\epsilon_{\mu}(p_1)~with~ \hat{s}=(p_3+p_4)^2.\end{equation}

The expressions for the amplitude of $Z(p_1,\epsilon_1)\to
\eta(p_3)\gamma(p_2,\epsilon_2)$ can be given by \bea
M&=&\frac{N_{cf}Q_f}{4\pi^2}(A+B)[\ea\cdot\eb(-2m_fC_\nu
p_2^{\nu}+\frac{m_f}{2}-m_f C_{\mu\nu}g^{\mu\nu}+m_fp_2\cdot
p_3C_0+m_f^3C_0)\nonumber\\&&+ p_2\cdot\eb(2m_f
C_\mu\ea^{\mu}-m_fp_3\cdot\ea C_0)+p_2\cdot\ea(2m_f
C_\nu\epsilon_2^{*\nu}-m_fp_3\cdot\eb
C_0)\nonumber\\&&-2m_fp_3\cdot\ea
C_\nu\epsilon_2^{*\nu}+4m_fC_{\mu\nu}\ea^{\mu}\epsilon_2^{*\nu}]+i\frac{N_{cf}Q_f}{4\pi^2}(C-D)m_f
\varepsilon_{\nu\alpha\mu\beta} \ea^\mu\epsilon_2^{*\nu}p_2^\alpha
p_3^\beta C_0, \eea where $N_{cf}$ and $Q_f$ denote the color factor
and the electric charge for the fermion $f$, respectively;
$A=\frac{1}{2}(g_L^Zg_R^\eta+g_R^Zg_L^\eta)$,
$B=\frac{1}{2}(g_L^Zg_L^\eta+g_R^Zg_R^\eta)$,
$C=\frac{1}{2}(g_L^Zg_R^\eta-g_R^Zg_L^\eta)$, and
$D=\frac{1}{2}(g_L^Zg_L^\eta-g_R^Zg_R^\eta)$. The parameters
$g_L^Z$, $g_R^Z$, $g_L^\eta$ and $g_R^\eta$ are respectively from
the couplings $i\bar{f}\gamma^{\mu}(g_L^Z P_L+g_R^Z P_R)fZ$ and
$\bar{f}(g_L^\eta P_L+g_R^\eta P_R)f\eta$ with
$P_{L,R}=\frac{1\mp\gamma^5}{2}$. $C_0$, $C_\mu$ and $C_{\mu\nu}$
are respectively the 3-point loop functions, and their dependence is
given by $C_{0,\mu,\mu\nu}\equiv
C_{0,\mu,\mu\nu}(p_2,-p_1,m_f,m_f,m_f).$

\end{document}